\documentstyle[12pt,epsf]{article}
\def\oop{{1\over{2\pi}}}
\def\intk{\int_{-k_0}^{k_0}}
\def\intl[#1]{\int_{-\Lambda_{#1}}^{\Lambda_{#1}}}
\def\SUN{SU({\it N})}
\def\atn{{\rm \, arctan\, }}
\def\mod{{\rm \, mod\, }}
\begin{document}

\title{
       Critical exponents of the degenerate Hubbard model
      }
\author{
        Holger Frahm\thanks{e-mail:
        {\tt frahm@kastor.itp.uni-hannover.de}}\\
        ~ \\
        {\normalsize Institut f\"ur Theoretische Physik,
                     Universit\"at Hannover}\\
        {\normalsize D-3000 Hannover 1, F.~R.~Germany}\\ ~ \\
\and
 {\large Andreas Schadschneider\thanks{e-mail:
        {\tt as@thp.uni-koeln.de}}~~\thanks{Work performed within the research
program of the Sonderforschungsbereich 341 (K\"oln-Aachen-J\"ulich)}} \\ ~ \\
        {\normalsize Institut f\"ur Theoretische Physik,
                     Universit\"at zu K\"oln} \\
        {\normalsize D-5000 K\"oln 41, F.~R.~Germany} \\ ~ \\
       }
\date{July 15, 1992}
\maketitle

% PACS-numbers: 05.30.Fk, 05.70.Jk, 71.30.+h
\centerline{Submitted to {\it J.\ Phys.\ A}}

\begin{abstract}
We study the critical behaviour of the \SUN{} generalization of the
one-dimensional Hubbard model with arbitrary degeneracy $N$. Using the
integrability of this model by Bethe Ansatz we are able to compute the
spectrum of the low-lying excitations in a large but finite box for
arbitrary values of the electron density and of the Coulomb
interaction. This information is used to determine the asymptotic
behaviour of correlation functions at zero temperature in the presence
of external fields lifting the degeneracy. The critical exponents
depend on the system parameters through a $N\times N$ dressed charge
matrix implying the relevance of the interaction of charge- and
spin-density waves.
\end{abstract}

\newpage

%%%%%%%%%%%%%%%%%%%%%%%%%%%%%%%%%%%%%%%%%%%%%%%%%%%%%%%%%%%%%%%%%%%%%%%%%%%
%
\section{Introduction}
%
%%%%%%%%%%%%%%%%%%%%%%%%%%%%%%%%%%%%%%%%%%%%%%%%%%%%%%%%%%%%%%%%%%%%%%%%%%%
The physics of highly correlated electron systems has long been the
subject of extensive studies in condensed matter physics.  Recently,
the non-Fermi liquid character of low-dimensional systems has
attracted renewed interest in one-dimensional realizations of these
systems where large quantum fluctuations lead to Luttinger liquid
behaviour \cite{hald8s}: the correlation functions decay as power-laws
at zero temperature, the exponents depending on the system parameters
such as electron density, magnetization (or applied magnetic field)
and strength of the interaction.

In this context exactly soluble models can provide a variety of new
insights, in particular when used together with the general results on
quantum critical behaviour in one spatial dimension as provided by the
theory of conformal invariance \cite{bpz}--\cite{bcn86}. Here the
universality class of the quantum system is completely determined by a
single dimensionless number---the so-called central charge $c$ of the
underlying Virasoro algebra. This number $c$ as well as the dimensions
of the operators present in the theory can be extracted from
analytical results for the spectrum of low-lying states in finite
geometries. In this language, Luttinger liquids correspond to a
central charge $c=1$, the dependence of the anomalous dimensions on
the system parameters is through a single dimensionless number---the
coupling constant of the corresponding Gaussian model.

The situation described above is the generic behaviour of 1D quantum
systems with a {\it single} critical degree of freedom, as realized in
spin-${1\over2}$ chains or systems of spinless fermions. The situation
becomes more complicated for systems where two or more massless
excitations are possible: the lack of Lorentz invariance (the
corresponding Fermi velocities differ in general) prevents the direct
application of the predictions of conformal field theory and the
interacting nature of the system complicates the factorization of the
problem into {\it independent} ones for each critical quasi-particle
mode. On the other hand the exact results on the finite-size scaling
of the low-lying energies available for Bethe Ansatz soluble models
suggests a resolution of this problem: the spectrum is that of a
multi-component Gaussian model. In analogy to the analysis of
conformal invariant theories the universality class of a system with
$N$ massless collective excitation modes is determined by $N$
dimensionless numbers $c_r$---reducing to the central charge in the
scalar case. The anomalous dimensions are functions of the system
parameters not through a single coupling constant but through a
$N\times N$ matrix of dimensionless numbers---the so-called dressed
charge matrix. This behaviour has been found in a large number of 1D
quantum systems, including certain integrable spin-chains with
$S>{1\over2}$ \cite{ikr89,fryu90}, and the Hubbard
\cite{woyn89,frko90,frko91} and $t-J$-models of correlated electrons,
the latter both at the integrable supersymmetric point $J/t=2$
\cite{kaya91} and away from integrability \cite{ogat91}.

Most of the integrable models in this list are solved by a hierarchy
of Bethe Ans\"atze \cite{yang67}: The first one introduces a set of
wavenumbers describing the phase of the wave function and determining
the spectrum, the others are necessary for the wavefunctions to show
the symmetry corresponding to a particular representation of the
Permutation group. This allows for the solution of certain systems
with various choices of internal degrees of freedom, e.g.\ an
SU(2)-spin in the Hubbard model.

In the present paper we study the critical properties of a
generalization of the Hubbard model, describing electrons carrying an
\SUN-spin index on the lattice. Unlike the case of the regular ($N=2$)
Hubbard model this model allows to study a Mott transition at a {\em
finite} value of the Coulomb interaction \cite{hald80,schl91,as91}.
Our paper is organized as follows: In the following section we shall
introduce the model and present a brief review of its Bethe Ansatz
solution and a qualitative discussion of its excitation spectrum. In
Section 3 the results for the finite-size corrections to the energies
of low-lying states as well as their relation to the critical
exponents are given in terms of the $N\times N$ dressed charge matrix.
In Section~4 the integral equations for this matrix are solved in the
zero-field case and the operator dimensions are computed as a function
of the electron density and the strength of the interaction. They are
shown to reflect the full \SUN-spin symmetry present in this case.
These results are applied to the computation of the critical exponents
for some correlation functions of interest.  Finally, in Section~5 we
consider states where the \SUN-symmetry of the ground state is broken
by magnetic fields coupling to the various flavours of the internal
degree of freedom.  In the limiting case of strong coupling we discuss
the dependence of the dressed charge matrix and the critical exponents
on these external fields.

%%%%%%%%%%%%%%%%%%%%%%%%%%%%%%%%%%%%%%%%%%%%%%%%%%%%%%%%%%%%%%%%%%%%%%
%% Date: Wed, 15 Jul 92 10:27:28 MET DST
%
%
\section{The Bethe-Ansatz solution of the degenerate Hubbard model}
%
%
%%%%%%%%%%%%%%%%%%%%%%%%%%%%%%%%%%%%%%%%%%%%%%%%%%%%%%%%%%%%%%%%%%%%%%
The Hamiltonian of the degenerate Hubbard model on a chain of length
$L$ is given by the following expression:
\begin{eqnarray}
   {\cal H}_N =&-&\sum_{j=1}^L \sum_{s=1}^N {\cal P} \left(
      c_{j+1,s}^\dagger c_{js} +c_{js}^\dagger c_{j+1,s}\right){\cal P}
      + 4u \sum_{j=1}^L\sum_{s,s'(s\neq s')} n_{js} n_{js'}\nonumber \\
   &+&\mu\sum_{j=1}^L\sum_{s=1}^Nn_{js}+\sum_{j=1}^L\sum_{s=1}^{N}h_sn_{js}
   \label{hamil}
\end{eqnarray}
The Fermi operator $c_{js}^\dagger $ ($c_{js}$) creates (annihilates) an
electron at site $j$ with spin index $s\in \{1,\ldots,N\}$ and
$n_{js}=c_{js}^\dagger c_{js}$ is the corresponding number operator.
The real parameters $h_s$ may be considered as generalized magnetic fields.
Note that these fields do not destroy the integrability of the model since
the numbers $N_s=\sum_{j=1}^Ln_{js}$ of particles with spin index $s$ are
conserved. ${\cal{P}}$ projects onto the subspace of states having at most two
electrons at each site. This projection is necessary in order to
render the Hamiltonian (\ref{hamil}) Bethe-Ansatz solvable
\cite{hald80,choy80,chha82,schl90} and
has no effect for $N=2$. The Bethe-Ansatz wave function which solves
the Schr\"odinger equation for (\ref{hamil}) for a total number $N_c$
of electrons is characterized by the momenta $k_j$ ($j=1,\ldots,N_c$)
and $N-1$ sets of rapidities $\lambda_\alpha^{(s)}$ ($s=1,\ldots,N-1$;
$\alpha=1,\ldots,M_s$).  Imposing periodic boundary conditions on the
wave function leads to the Bethe-Ansatz equations
%\footnote{Recently, Schlottmann \cite{schl92} raised the question whether
%the eqs.\ (\ref{bae}) are the Bethe-Ansatz equations of (\ref{hamil}) or a
%more complicated generalized Hubbard model.}
%
\begin{eqnarray}
   Lk_j &=& 2\pi I_j-\sum_{\beta=1}^{M_1}2\atn
               \left({\sin k_j -\lambda_\beta^{(1)} \over u}\right),
   \nonumber \\
   \sum_{\beta=1}^{M_{s-1}}2\atn
            \left({\lambda_\alpha^{(s)} -\lambda_\beta^{(s-1)} \over u}\right)
         &+& \sum_{\beta=1}^{M_{s+1}}2\atn \left({\lambda_\alpha^{(s)} -
         \lambda_\beta^{(s+1)} \over u}\right) \nonumber \\
   &=& 2\pi J_\alpha^{(s)} + \sum_{\beta=1}^{M_{s}}2\atn
      \left({\lambda_\alpha^{(s)} -\lambda_\beta^{(s)} \over 2u}\right).
   \label{bae}
\end{eqnarray}
Here we have set $M_0=N_c$, $M_N=0$ and $\lambda_j^{(0)}=\sin k_j$.

The quantum numbers $I_j$ and $J_\alpha^{(s)}$ are integer or
half-integer depending on the parity of the numbers $N_c$,
$M_\alpha^{(s)}$:
\begin{equation}
   I_j={M_1\over 2} \mod 1, \quad
   J_\alpha^{(s)}= {M_s-M_{s-1}-M_{s+1}+1\over 2} \mod 1.
   \label{parity}
\end{equation}
Energy and momentum of the model in a state corresponding to a
solution of (\ref{bae}) are completely determined by the momenta
$k_j$:
\begin{eqnarray}
   E&=&-2\sum_{j=1}^{N_c}\cos k_j +\mu N_c -\sum_{s=1}^{N}h_sN_s,\nonumber \\
   P&=&\sum_{j=1}^{N_c} k_j ={2\pi \over L} \left(\sum_{j=1}^{N_c}I_j+
   \sum_{s=1}^{N-1}\sum_{\alpha=1}^{M_s} J_\alpha^{(s)}\right),
   \label{eigenvals}
\end{eqnarray}
where $N_s$ denotes the total number of electrons with orbital index
$s$.  In the thermodynamic limit ($L\to\infty$, with $N_c/L$, $M_s/L$
kept constant) the eqs.\ (\ref{bae}) corresponding to the ground state
of (\ref{hamil}) can be transformed into a set of coupled integral
equations for the densities $\rho_c(k)$ and $\rho_s(\lambda)$ of the
parameters $k_j$ and $\lambda_\alpha^{(s)}$, respectively:
\begin{eqnarray}
   \rho_c(k)&=&\oop+{\cos k \over 2\pi}\intl[1] d\lambda \, K_1(\sin k-\lambda)
   \rho_{1}(\lambda)\, ,
   \nonumber \\
   \rho_{1}(\lambda)&=& \oop\intk dk \, K_1(\lambda -\sin k)\rho_c(k)
    -\oop\intl[1] d\mu\, K_2(\lambda -\mu)\rho_{1}(\mu)
        \nonumber \\
    &&+\oop\intl[2] d\mu \, K_1(\lambda - \mu)\rho_{2}(\mu)\, ,
   \label{baerho} \\
   \rho_{s}(\lambda)&=&\oop\intl[s-1] d\mu\, K_1(\lambda -\mu)\rho_{s-1}(\mu)
     -\oop\intl[s] d\mu \, K_2 (\lambda -\mu){\rho_{s}}(\mu)   \nonumber \\
    &&+\intl[s+1]d\mu\, K_1(\lambda -\mu)\rho_{s+1}(\mu),
    \quad\quad\quad (s=2,\ldots,N-1)\nonumber
\end{eqnarray}
with $\Lambda_N=0$.
The kernels $K_{1,2}(x)$ of these eqs.\ (\ref{baerho}) are
given by
\begin{equation}
K_1(x)={2u \over x^2+u^2}, \quad K_2(x)={4u \over x^2+(2u)^2}.
\label{ikernels}
\end{equation}
The values of the parameters $k_0$ and
$\Lambda_1,\ldots,\Lambda_{N-1}$ are determined through the
normalizations
\begin{eqnarray}
    n_c&=&\intk dk\, \rho_c(k)\, ,\nonumber \\
%   n_1&=&n_c-\intl[1]d\lambda \, \rho_{1}(\lambda)\, ,
    n_s&=&\intl[s-1]d\lambda\, \rho_{s-1} (\lambda)
         -\intl[s]d\lambda \, \rho_{s} (\lambda)\, ,\quad\quad\quad
    (s=1,\ldots,N-1)   \label{densities}
\end{eqnarray}
where $n_c=N_c/L$ is the total density of electrons and
$n_s=N_s/L=(M_{s-1}-M_s) /L$ is the density of electrons with index
$s$. Furthermore we have set $\rho_0\equiv\rho_c$ and $\Lambda_0=
k_0$.\footnote{%
In the following we will adopt the convention that an index '0' stands
for 'c'.}

The ground state energy per lattice site is
\begin{equation}
  \epsilon_\infty=\intk dk\, (\mu+h_1-2\cos k)\rho_c(k)+\sum_{s=1}^{N-1}
   \left( h_{s+1}-h_s\right)\intl[s]d\lambda \, \rho_s(\lambda)
\label{eground}
\end{equation}
which may alternatively be expressed in terms of the dressed energy
\begin{equation}
  \epsilon_\infty=\oop\intk dk \,\varepsilon_c(k).
\label{eground2}
\end{equation}
Here $\varepsilon_c(k)$ is the solution of the system of coupled integral
equations
\begin{eqnarray}
   \varepsilon_c(k)&=&\varepsilon_c^{(0)}(k)+\oop\intl[1] d\lambda \,
     K_1(\sin k-\lambda)\varepsilon_{1}(\lambda),   \nonumber \\
    \varepsilon_{1}(\lambda)&=&\varepsilon_1^{(0)}(\lambda)+
     \oop\intk dk \cos k \, K_1(\lambda - \sin k) \varepsilon_c(k)-
     \oop\intl[1] d\mu K_2(\lambda -\mu )\varepsilon_{1}(\mu)\nonumber\\
    &&+\oop\intl[2] d\mu K_1(\lambda -\mu )\varepsilon_{2}(\mu),
  \label{dressen} \\
    \varepsilon_{s}(\lambda)&=&\varepsilon_s^{(0)}(\lambda)+
     \oop\intl[s-1] d\mu \,  K_1(\lambda - \mu) \varepsilon_{s-1}(\mu)-
     \oop\intl[s] d\mu K_2(\lambda -\mu )\varepsilon_{s}(\mu)\nonumber\\
     &&+\oop\intl[s+1] d\mu K_1(\lambda -\mu )\varepsilon_{s+1}(\mu)
     \quad\quad\quad (s=2,\ldots,N-1).
   \nonumber
\end{eqnarray}
The bare energies are from (\ref{eigenvals})
\begin{equation}
   \varepsilon_c^{(0)}(k)=\mu +h_1-2\cos k,\quad \varepsilon_s^{(0)}(\lambda)=
    h_{s+1}-h_s.
\label{bareener}
\end{equation}
The dressed energies (\ref{dressen}) obey the conditions
\begin{equation}
   \varepsilon_c(k_0)=0,\quad \varepsilon_s(\Lambda_s)=0.
\label{fermivals}
\end{equation}

The ground state at half-filling ($n_c=1$) and with vanishing fields $h_s$
shows for $N>2$ an interesting behavior which has not been found in the case of
the standard Hubbard model ($N=2$). This has been noticed independently by
Schlottmann \cite{schl91} and one of the authors \cite{as91}. For $u>u_c$ one
finds $k_0=\pi$ and for $u<u_c$ one has $k_0<\pi$.  Here the critical value
$u_c$ is determined through the implicit equation
\begin{equation}
   \int_{-\pi}^\pi dk\, G_N(\sin k;u_c) =2\pi
   \label{ucrit}
\end{equation}
where $G_N(x;u)$ in terms of the Digamma-function $\psi(x)$ is given by
\begin{equation}
G_N(x;u)={1 \over Nu}\,{\rm Re}\left[ \psi \left(1+i{x\over 2Nu}\right)
-\psi \left({1\over N}+i{x\over 2Nu}\right)\right]\, .
\label{effkernel}
\end{equation}
For $N=2$ we have $u_c=0$ (see Fig.\ 1) as already shown by Lieb and Wu
\cite{lw68}.

The excitation spectrum in zero-fields has also been studied in \cite{as91}
using an extension of the method developed in \cite{ksz90,sz91} and in
\cite{schl91}. One finds $N-1$ gapless spin excitations with soft modes with
wave numbers $2sP_F$ ($s=1,\ldots,N-1$, $P_F={\pi\over N}n_c$)
and so-called particle-hole excitations.\footnote{%
For a discussion of the zero-field excitation spectrum for $N=2$, see e.\ g.\
\cite{ksz90} and references therein.}
These also are gapless and do exist only for $k_0<\pi$, i.\ e.\ for
$n_c<1$ or $n_c=1$ and $u<u_c$. In \cite{as91} also excitations corresponding
to doubly occupied sites have been studied. These are described by complex
momenta $k^\pm$ satisfying $\sin k^\pm=\lambda \pm iu$ and have a finite gap
(at least for $n_c=1$ and $u>u_c$).

The special structure of the ground state for $n_c=1$ leads to interesting
properties of the model. For $u>u_c$ the only possible charge carrying
excitations are those involving complex momenta. As these excitations
have a gap the system is in an insulating state. For $u<u_c$ particle-hole
excitations become possible. These excitations may carry a current and so the
system is in a metallic phase. This shows the existence of Mott-transition
at the critical value $u_c$ of the Coulomb repulsion $u$. The transition
is also reflected in the behavior of other physical quantities, e.\ g.\ the
charge susceptibility $\chi_c$ and the Fermi velocity $v_c$ \cite{schl91,as91}.
As the gap to the excitations with double occupations does not vanish in the
limit $u\searrow u_c$ the value of the gap to the charge carrying shows a
discontinuity at $u=u_c$. We may thus say that the transition is of
'first order'.

In the limit $u\to \infty$ at $n_c=1$ the model (\ref{hamil}) becomes
equivalent to the \SUN\ Heisenberg chain \cite{su75}. This equivalence
generalizes the well-known relation between the regular ($N=2$) Hubbard model
and the Heisenberg antiferromagnet.

Recently, Schlottmann \cite{schl92} showed that in the continuum limit the
particles interact via a potential of the form $1/\sinh^2r$ where $r$ is the
distance between the particles involved in some properly chosen units. This
reflects the nonlocal character of the interaction as introduced by
the projectors ${\cal P}$ in the kinetic terms of (\ref{hamil}) for
$N>2$.

%%%%%%%%%%%%%%%%%%%%%%%%%%%%%%%%%%%%%%%%%%%%%%%%%%%%%%%%%%%%%%%%%%%%%%%%%%%%%
%
%
\section{Finite-size corrections and conformal properties}
%
%
%%%%%%%%%%%%%%%%%%%%%%%%%%%%%%%%%%%%%%%%%%%%%%%%%%%%%%%%%%%%%%%%%%%%%%%%%%%%%
As shown in the preceding section the degenerate Hubbard model supports gapless
excitations in general. Thus we may apply the concepts of conformal field
theory to determine the asymptotic behavior of the correlation functions,
e.\ g.\ the critical exponents.

First we calculate exactly the finite-size corrections to the ground state
energy and the energies of the excited states. This can be done by a
straightforward extension of the calculation for the case $N=2$ \cite{woyn89}.
The results can be expressed in terms of the $N\times N$-dressed charge
matrix
\begin{equation}
  Z=\pmatrix{
     \xi_{cc}(k_0) & \xi_{c1}(\Lambda_1) & \cdots & \xi_{c,N-1}(\Lambda_{N-1})
	   \cr
     \xi_{1c}(k_0) & \xi_{11}(\Lambda_1) & \cdots & \xi_{1,N-1}(\Lambda_{N-1})
           \cr
     \vdots        & \vdots              &        & \vdots   \cr
     \xi_{N-1,c}(k_0) & \xi_{N-1,1}(\Lambda_1) & \cdots & \xi_{N-1,N-1}
           (\Lambda_{N-1}) \cr}.
     \label{defz}
\end{equation}
The elements of $Z$ can be obtained from the dressed-charge functions
$\xi_{rs}(\lambda)$ which obey a system of coupled integral
equations similar to (\ref{dressen}). For $r=c,1,\ldots,N-1$ we have
\begin{eqnarray}
   \xi_{r c}(k)&=&\delta_{r c}+\oop\intl[1] d\lambda \cos k \,
     K_1(\sin k-\lambda)\xi_{r 1}(\lambda),   \nonumber \\
    \xi_{r 1}(\lambda)&=&\delta_{r 1} +
     \oop\intk dk \cos k \, K_1(\lambda - \sin k) \xi_{r c}(k)-
     \oop\intl[1] d\mu K_2(\lambda -\mu )\xi_{r 1}(\mu)\nonumber\\
    &&+\oop\intl[2] d\mu K_1(\lambda -\mu )\xi_{r 2}(\mu),
  \label{dressch} \\
    \xi_{rs}(\lambda)&=&\delta_{rs}+
     \oop\intl[s-1] d\mu \,  K_1(\lambda - \mu) \xi_{r,s-1}(\mu)-
     \oop\intl[s] d\mu K_2(\lambda -\mu )\xi_{rs}(\mu)\nonumber\\
     &&+\oop\intl[s+1] d\mu K_1(\lambda -\mu )\xi_{r,s+1}(\mu)
     \qquad\quad\quad (s = 2,\ldots,N-1).
   \nonumber
\end{eqnarray}
The finite-size scaling behavior of the ground-state energy is found to be
\begin{equation}
    E_0-L\epsilon_\infty=-{\pi \over 6L} \sum_{s=0}^{N-1}v_s
	\label{groundfss}
\end{equation}
with the Fermi velocities of charge and spin excitations
\begin{equation}
   v_0\equiv v_c={1\over 2\pi\rho_c(k_0)}\,\varepsilon_c'(k_0), \quad
   v_s={1\over 2\pi\rho_s(\Lambda_s)}\, \varepsilon_s'(\Lambda_s)
        \qquad\quad (s=1,\ldots,N-1).
	\label{fermivels}
\end{equation}
Energies and momenta of the excitations scale as
\begin{eqnarray}
   E(\Delta{\bf M},{\bf D})-E_0 &=& {2\pi \over L}\left[ {1\over 4}\Delta
     {\bf M}^T\left( Z^{-1}\right)^T V Z^{-1} \Delta{\bf M} + {\bf D}^TZVZ^T
     {\bf D}
%     \nonumber\\
     + \sum_{s=0}^{N-1}v_s(N_s^++N_s^-)\right], \nonumber \\
   P(\Delta{\bf M},{\bf D})-P_0 &=& {2\pi\over L}\left[ \Delta{\bf M}^T\cdot
     {\bf D} + \sum_{s=0}^{N-1}(N_s^+-N_s^-)\right]
     + 2\sum_{s=0}^{N-1}\sum_{r=0}^{s}D_r P_{F,s+1}, \label{excitfss}\\
   V&=&{\rm diag}(v_c,v_1,\ldots,v_{N-1}).\nonumber
\end{eqnarray}
Here $N_c^\pm$, $N_s^\pm$ are positive integers and $\Delta \bf M$ and $\bf D$
are vectors characterizing the excited state under consideration. $P_{F,s}$
are the Fermi momenta for electrons with spin index $s$. For the
ground state in the thermodynamic limit  we have $\Delta M_s =0$, $D_s=0$
($s=c,1,\ldots,N-1$). For an excited state $\Delta\bf M$ has integer components
denoting the change of the total number of electrons and the number of
electrons
with index $s$ with respect to the ground state. $D_s$ are integer or
half-odd integer depending on the parities of the $\Delta M_s$. Due to
(\ref{parity}) we have
\begin{eqnarray}
   D_c &=& {\Delta N_c + \Delta M_1 \over 2}\, \mod 1, \nonumber \\
   D_s &=& {\Delta M_{s-1} + \Delta M_{s+1} \over 2}\,\mod 1 \qquad\quad
     (s=1,\ldots,N-1) \label{parityd}
\end{eqnarray}
with $\Delta M_0=\Delta N_c$ and $\Delta M_N=0$.

In general, all velocities $v_s$ are different. In this case the results
(\ref{groundfss}) and (\ref{excitfss}) may be interpreted in terms of a
semidirect product of $N$ independent Virasoro algebras.\footnote{See
\cite{frko90} and references therein for a more detailed discussion of this
point.} All these Virasoro algebras have central charge $c_s=1$. For
vanishing fields $h_s$ all magnon velocities $v_1,\ldots,v_{N-1}$ are
equal \cite{schl91,as91}  and we have a semidirect product of a $c=1$ Gaussian
theory -- reflecting the U(1) symmetry of the charge sector -- and a $c=N-1$
Wess-Zumino-Witten theory -- reflecting the \SUN-symmetry of the spin sector.

Comparing (\ref{excitfss}) with the predictions of the conformal field
theory \cite{card86,bcn86}
\begin{eqnarray}
   E(\Delta{\bf M},{\bf D})-E_0 &=& {2\pi \over L}\sum_{s=0}^{N-1}v_s
      (\Delta_s^++\Delta_s^-), \nonumber \\
   P(\Delta{\bf M},{\bf D})-P_0 &=& {2\pi\over L}\sum_{s=0}^{N-1}(\Delta_s^+-
      \Delta_s^-)+2\sum_{s=0}^{N-1}\sum_{r=0}^{s}D_r P_{F,s+1},
      \label{cftpred}
\end{eqnarray}
one obtains expressions for the conformal dimensions $\Delta_s^{\pm}$ of the
primary fields in terms of the dressed charge matrix. Requiring that all
dimensions are positive we find
\begin{equation}
    2\Delta_{s}^\pm=\left(\left(Z^T{\bf D}\right)_s\pm {1\over 2}\left(
         Z^{-1}\Delta{\bf M}\right)_s\right)^2+2N_s^\pm \label{dimens}
\end{equation}
which, in general, depend on the system parameters. In the following section
we will show that for vanishing fields the $\Delta_s^\pm$ ($s=1,\ldots,N-1$)
are functions of the components of $\Delta{\bf M}$ and ${\bf D}$ only,
whereas $\Delta_c^\pm$ depends on the strength $u$ of the Coulomb repulsion
and the density $n_c$ of electrons.

We now make further use of the results of conformal field theory to write
down the correlation functions for primary fields as
\begin{equation}
\langle \phi_{\Delta^\pm}(x,t)\phi_{\Delta^\pm}(0,0)\rangle=
\prod_{s=0}^{N-1}{\exp\left(-2i\sum_{r=0}^{s}D_r P_{F,s+1}x\right) \over
(x-iv_st)^{2\Delta_s^+}(x+iv_st)^{2\Delta_s^-}}.\label{prikorr}
\end{equation}
The correlation functions of the physical fields consist of a sum of
terms (\ref{prikorr}). In the following we will study correlators of the
form $\langle {\cal O}_j(t){\cal O}^\dagger_0(0)\rangle$ where ${\cal O}$ is
given in terms of $c$ and $c^\dagger$. To find the asymptotic behavior of
the correlator one has to expand ${\cal O}$ in terms of the conformal fields.
This is not possible in general, but the explicit form of ${\cal O}$ allows for
an identification of the quantum numbers  $M_c,M_1,\ldots,M_{N-1}$ of the
intermediate states. Therefore the leading term in the asymptotic expansion
of $\langle {\cal O}_j(t){\cal O}^\dagger_0(0)\rangle$ can be obtained from
(\ref{dimens}) through minimizing with respect to  the $D_s$ satisfying
(\ref{parityd}).

%%%%%%%%%%%%%%%%%%%%%%%%%%%%%%%%%%%%%%%%%%%%%%%%%%%%%%%%%%%%%%%%%%%%%
%
%
\section{Critical exponents for vanishing fields}
%
%
%%%%%%%%%%%%%%%%%%%%%%%%%%%%%%%%%%%%%%%%%%%%%%%%%%%%%%%%%%%%%%%%%%%%%%
In the absence of magnetic fields it is easily seen that
$\Lambda_r=\infty$ for all $r$.  This allows to eliminate the
$\lambda$-dependent quantities from the Bethe Ansatz integral
equations by Fourier transformation.  From Eq.~(\ref{dressch}) we
obtain for the $k$-dependent entries of the dressed charge matrix
\begin{equation}
   \xi_{rc}(z) = {{N-r}\over N}
               + \oop \int_{-z_0}^{z_0} dy \xi_{rc}(y) G_N(z-y;1).
   \label{dressch0}
\end{equation}
where we have introduced a new variable $z=\sin{}k/u$.

The solution for $z_c=\xi_{cc}(z_0)$ is obtained by iteration for
small values of $z_0$ where we obtain
\begin{equation}
       z_c \simeq 1 + \frac{G_N(0;1)}{\pi}z_0
           = 1 - \frac{\gamma+\psi({1\over N})}{N\pi} z_0
       \qquad {\rm for~}z_0\ll 1.
\end{equation}
For $z_0\gg N$ a perturbative scheme \cite{yaya66} based on the
Wiener-Hopf method can be applied, giving
\begin{equation}
       z_c \simeq \sqrt{N} \left( 1-\frac{N-1}{2\pi z_0} \right)
       \qquad {\rm for~}z_0\gg N.
\end{equation}
For intermediate values of $z_0$ the integral equation
(\ref{dressch0}) is easily solved numerically. The dependence of $z_c$
on the density and the strength of the Coulomb interaction is shown in
Fig.~\ref{fig:zcontours} for some values of $N$.  In addition to the
interpretation of the $z_c$ as measure of the reordering of the
Fermi-sea due to the interaction when an electron is added there
exists a direct relation to physical observables: It can be expressed
as
\begin{equation}
   z_c^2 = \pi v_c n_c^2 \kappa
\end{equation}
in terms of the compressibility $\kappa=-(1/L)\partial{}L/\partial{}p$
of the electron gas ($p$ being the pressure).

As in the regular Hubbard model ($N=2$) \cite{woyn89} one can employ
the Wiener-Hopf method to compute the remaining elements of the
dressed charge matrix (\ref{defz}) yielding:
\begin{equation}
   Z = \left( \begin{array}{ccccc}
                 z_c & \begin{array}{cccc} 0&0&\cdots&0\end{array} \\ [8pt]
              \begin{array}{c}
                  {N-1\over N} z_c \\ [8pt]
                  {N-2\over N} z_c \\ [8pt]
                      \vdots       \\ [8pt]
                    {1\over N} z_c
              \end{array} & {\cal Z}_N
               \end{array}
        \right)  \label{zmatrix}
\end{equation}
Note that the symmetric $(N-1)\times(N-1)$-block ${\cal Z}_N$ of this
matrix is completely determined by the \SUN-spin-symmetry of the
system: this symmetry is manifest in the kernel of the integral
equations and allows for the reduction of the corresponding {\it
matrix} Wiener-Hopf problem to $N-1$ {\it scalar} ones
\cite{vega88,suzu88}. The latter are soluble by quadratures and one
obtains closed expressions for the matrix elements of ${\cal Z}_N^{-1}$:
\begin{equation}
   \left({\cal Z}_N^{-1}\right)_{rs} =
      - {2\over N} {\sin{}{\pi\over{2N}}}
              {\sqrt{1-z_r^2}\sqrt{1-z_s^2} \over
              {z_r^2+z_s^2-2\cos{}{\pi\over{2N}} z_r z_s
                    - \sin^2{\pi\over{2N}}}},
          \qquad z_n = \cos{\pi{}n\over N}.
   \label{znmatrix}
\end{equation}
The square of ${\cal Z}_N^{-1}$ is the Cartan matrix for the Lie
algebra \SUN:
\[
     \left({\cal Z}_N^{-2}\right)_{rs} = \left({\cal C}_N\right)_{rs}
        =2\delta_{r,s} - \delta_{r,s+1} - \delta_{r+1,s}.
\]
Using the properties of the dressed charge matrix we find for the
critical exponents
\begin{eqnarray}
   2\Delta_c^{\pm}&=&\left( z_c\sum_{s=0}^{N-1}{N-s\over N}D_s
                           \pm{1\over 2z_c} \Delta N_c\right)^2 +2N_c^\pm
  \nonumber \\ \label{dimsh0} \\
   2\Delta_r^{\pm}&=&\left( \sum_{s=1}^{N-1}({\cal Z}_N)_{sr}D_s
                      \pm{1\over 2} \sum_{s=1}^{N-1}
                       \left({\cal Z}_N^{-1}\right)_{rs}
                        \left(\Delta M_s
         -{N-s\over N} \Delta N_c\right)\right)^2 +2N_r^\pm
  \nonumber
\end{eqnarray}
For vanishing magnetic field the magnon velocities $v_r$
($r=1,\ldots,N-1$) are identical, hence only the quantities
\begin{eqnarray}
   2\Delta_\sigma^{\pm} = 2 \sum_{r=1}^{N-1} \Delta_r^{\pm}
        = {1\over 4}{\Delta\bf N}_\sigma^T {\cal C}_N {\Delta\bf N}_\sigma +
                           {\bf D}_\sigma^T {\cal C}_N^{-1} {\bf D}_\sigma \pm
                           {\Delta\bf N}_\sigma^T \cdot {\bf D}_\sigma
          + 2 \sum_{r=1}^{N-1} N_r^{\pm} ,
   \nonumber \\
   {\rm where~} \left({\Delta\bf N}_\sigma\right)_r
                   = \Delta M_r -{N-r\over N} \Delta N_c, \quad
                \left({\bf D}_\sigma\right)_r = D_r, \quad r=1,\ldots,N-1
\end{eqnarray}
appear in the exponents describing the asymptotic behaviour of the
correlation functions.  Note that they are independent of the quantity
$z_c$ incorporating the dependence of the anomalous dimensions on
electron density and strength of the interaction.  They are completely
determined by the \SUN{}-symmetry of the zero field ground state and,
in fact, of the same form as the exponents characterizing the
\SUN-symmetric critical vertex models and spin-chains
\cite{vega88,suzu88} (the difference being the possibility
of fractional values for the elements of ${\Delta\bf N}_\sigma$).

Now we are able to study the asymptotic behaviour of correlation
functions of interest:\\
For the field-field correlation function we have $\Delta{}N_c=1$
and $\Delta{\bf M}=(1,\ldots,1_k,0,\ldots)$. The corresponding values
of {\bf D} can be read off from Eq.~(\ref{parityd}), the contribution
at wavenumber $k={P}_F=(\pi/N)n_c$ arises from the choice
$D_{r}={1\over2}(\delta_{rs}-\delta_{r,s+1})$ for any $s$ giving
\begin{equation}
   2\Delta_c^\pm = \left({z_c\over{2N}} \pm {1\over{2z_c}} \right)^2, \qquad
   2\Delta_\sigma^+ = \left({1} - {1\over{N}}\right), \qquad
   2\Delta_\sigma^- = 0.
\end{equation}
Hence, the singularity of the momentum distribution function at the
Fermi-point
\begin{equation}
   \langle c_{k,s} c_{k,s}^\dagger \rangle \sim |k-{P_F}|^\alpha
\end{equation}
is characterized by the exponent
\begin{equation}
   \alpha= {\theta\over(2N)^2} + {1\over\theta} -{1\over N}
   \label{alpha_h0}
\end{equation}
where $\theta=2z_c^2$ varies between 2 and $2N$ as $u$ decreases from
$\infty$ to 0.  Hence,
\begin{equation}
   0 = \alpha(u\to0) < \alpha < \alpha(u\to \infty)
     = {1\over2}-{1\over N}+{1\over{2N^2}}
\end{equation}
showing the Luttinger liquid character of the degenerate Hubbard model
at nonzero interaction. Numerical data for the exponent $\alpha$ are
presented in Fig.~\ref{fig:alpha}.

For the density-density correlation function $\langle n_s(x,t) n_s(0)
\rangle - \langle n \rangle^2$ we find contributions at wavenumbers
$k=2m{P}_F$ ($m=1,\ldots,N$) with exponents
\begin{equation}
   2\Delta_c^\pm = 2\left({m\over 2N}\right)^2 \theta, \qquad
   2\Delta_\sigma^\pm = {{m(N-m)}\over N}
\end{equation}
arising from the choice $\Delta N_c=\Delta M_r=0$ and
$D_r=\delta_{r,N-m}$.  In addition, there are $k=0$-terms decaying as
$x^{-2}$ asymptotically. They are generated by the marginally relevant
secondary operators in the conformal family of the unit operator
(i.e.\ $\Delta N_c=\Delta M_r=0$ and $D_c=D_r=0$ but $N_c^\pm$ or one
of the $N_r^\pm$ in Eq.~(\ref{dimsh0}) equal to 1).

{}From the discussion in the preceeding section it is clear that the
above statements are valid for any value of the coupling constant and
for any filling with $n_c<1$. For $n_c=1$ two cases have to be
discussed separately:

(i) For $n_c=1$ and $u\le u_c$ the system is in a metallic phase
(described by $k_0=k_0(u)\le\pi$ in the Bethe Ansatz equations).  The
critical exponents of the system are given by (\ref{dimsh0}) with
$z_c$ being a function of the Coulomb interaction through
(\ref{dressch0}). The number $z_{c}$ decreases from $\sqrt{N}$ to 1 as
the strength of the Coulomb interaction is varied from 0 to $u_c$.  It
is possible to expand the integral equations for the density
$\rho_c(k)$ in the neighbourhood of the Mott transition $u\le u_c$ to
determine $k_0(u)$ from (\ref{densities}).  For $N\to\infty$ the
resulting expressions simplify, giving $u_c=\frac{1}{2}\sqrt{3}$
\cite{hald80} and
\begin{equation}
   z_c\approx 1 + {{4\sqrt{14}}\over{7 \pi}} \sqrt{1-{u\over u_c}}
              \; + \ldots
\end{equation}
The same square-root singularity in the Coulomb coupling $u$---but
with a different numerical prefactor---is found near $u=u_c$ for finite
$N$. Through (\ref{dimsh0}) it also appears in the critical exponents
near the Mott transition, e.g.\ the exponent $\alpha$ of the momentum
distribution (\ref{alpha_h0}) varies like
\begin{equation}
   \alpha \approx {1\over2} -
   {{4\sqrt{14}}\over{7 \pi}} \sqrt{1-{u\over u_c}} \; + \ldots
\end{equation}
as $u$ approaches $u_c$ from below (for $N\to\infty$).

(ii) For $u>u_c$ with one particle per site the system is in an
insulating state, hence charge carrying excitations develop a gap.
Excitations in the spin degrees of freedom, however, continue to be
massless at zero temperature.  The critical properties of this state
can be described along the lines of the discussion above:
For $n_c=1$ and $u>u_c$ we have $k_0=\pi$.  Hence, the $k$-dependent
quantities disappear from Eq.~(\ref{dressch}) leaving a system of
$N-1$ coupled integral equations for the spin components of the
dressed charge matrix.  The expression for the conformal dimensions
are of the form (\ref{dimens}) and depend on the applied magnetic
fields. The correlation functions for states corresponding to critical
excitations are given by (\ref{prikorr}). These are exactly those
found in the \SUN-generalization of the Heisenberg spin chain (see
e.g.\ \cite{ikr89}). Note, that (\ref{parityd}) implies that the
momentum of the intermediate state is shifted by $\pi$ for states with
odd $\Delta{}M_1$.

%%%%%%%%%%%%%%%%%%%%%%%%%%%%%%%%%%%%%%%%%%%%%%%%%%%%%%%%%%%%%%%%%%%%%%
%
%
\section{Magnetic field effects in the strong coupling regime}
%
%
%%%%%%%%%%%%%%%%%%%%%%%%%%%%%%%%%%%%%%%%%%%%%%%%%%%%%%%%%%%%%%%%%%%%%%
Nonzero magnetic fields $h_s$ in (\ref{bareener}) lead to finite
values of the parameters $\Lambda_s$ through (\ref{fermivals}). This
effect in turn leads to a general dependence of the elements of the
dressed charge matrix on the system parameters $u$, $n_c$ and all of
the fields.  Only at and beyond certain critical values of the fields
where one or more bands are completely depleted the integral equations
simplify to some extent so that analytical results may become
available (for a discussion of the $N=2$ case see \cite{frko90}). A
more detailed study of the magnetic field dependence of the dressed
charge matrix becomes possible in the strong coupling limit
\cite{frko91}. As in the SU(2) Hubbard model the Bethe Ansatz integral
equations (\ref{baerho}), (\ref{dressen}) and (\ref{dressch})
describing the system simplify in the limit $u \to\infty$ (see also
\cite{schl89}).  In this limit the $k$-dependent quantities can be
eliminated which allows to study the effect of magnetic fields on the
critical exponents in more detail. Upon rescaling of the variables
$\lambda/u \to \tilde{\lambda} \equiv\lambda$ one obtains to leading order
\begin{eqnarray}
   \varepsilon_r(\lambda) &=& e_r^{(0)}(\lambda)
         - \oop \intl[r] d\mu \tilde{K}_2(\lambda-\mu) \varepsilon_{r}(\mu)
     \nonumber \\
     && + \oop \intl[r-1] d\mu \tilde{K}_1(\lambda-\mu) \varepsilon_{r-1}(\mu)
        + \oop \intl[r+1] d\mu \tilde{K}_1(\lambda-\mu) \varepsilon_{r+1}(\mu)
   \label{dressen_sc}
\end{eqnarray}
($r=1,\ldots,N-1$) where $\Lambda_0=\Lambda_N\equiv0$ and
\begin{equation}
   e_r^{(0)} = \varepsilon_r^{(0)} - {{\sin{}2k_0-2k_0}\over
                            {\pi{}u(1+\lambda^2)}} \delta_{r,1}
                \equiv \varepsilon_r^{(0)} - {h_c\over{1+\lambda^2}}
                     \delta_{r,1}.
   \label{bareen_sc}
\end{equation}
The integration kernels $\tilde{K}_i$ are obtained from
Eq.~(\ref{ikernels}) by setting $u=1$.
The dressed energy of the charged excitations is
\begin{equation}
   \varepsilon_c(k) = -2 \left(\cos{}k - \cos{}k_0 \right).
\end{equation}

Similarly, one finds reduced integral equations for the
spin-components of the density
\begin{eqnarray}
   \rho_r(\lambda) &=& {n_c\over \pi(1+\lambda^2)} \delta_{r,1}
           - \oop \intl[r] d\mu \tilde{K}_2(\lambda-\mu) \rho_{r}(\mu)
     \nonumber \\
        && + \oop \intl[r-1] d\mu \tilde{K}_1(\lambda-\mu) \rho_{r-1}(\mu)
           + \oop \intl[r+1] d\mu \tilde{K}_1(\lambda-\mu) \rho_{r+1}(\mu),
     \label{dressrho_sc}
\end{eqnarray}
and of the dressed charge matrix ($r,s = 1,\ldots,N-1$):
\begin{eqnarray}
    \xi_{sr}(\lambda)&=&\delta_{sr}
        - \oop\intl[s] d\mu \tilde{K}_2(\lambda -\mu )\xi_{sr}(\mu)\nonumber\\
     && + \oop\intl[s-1] d\mu \,  \tilde{K}_1(\lambda - \mu) \xi_{s,r-1}(\mu)
        + \oop\intl[s+1] d\mu \tilde{K}_1(\lambda -\mu )\xi_{s,r+1}(\mu)
     \label{dressch_sc}
\end{eqnarray}
The other elements of the dressed charge matrix are found to be
\begin{eqnarray}
   Z_{cc} = 1, \qquad Z_{cr}=0, \quad r=1,\ldots,N-1 \nonumber \\
   Z_{rc} = \oop \intl[1] d\lambda \tilde{K}_1(\lambda) \xi_{r1}(\lambda).
\end{eqnarray}
The expression for $Z_{rc}$ can be rewritten using the symmetry of the
kernel in (\ref{dressch_sc}) to obtain a simple relation to the
densities of electrons with \SUN{} index $s$:
\begin{equation}
   Z_{rc} = {1\over n_c} \intl[r] d\lambda \rho_r(\lambda)
          = 1 - \sum_{s=1}^{r} {n_s \over n_c}
   \label{zzch_sc}
\end{equation}
(since $n_s=n_c/N$ for $h_s\equiv 0$ this reproduces the corresponding
entries in Eq.~(\ref{zmatrix}) in the $u\to\infty$ limit).

For small magnetic fields (corresponding to large but finite values of
the $\Lambda_r$) one can employ the Wiener-Hopf method to the integral
equations (\ref{dressen_sc}) for the dressed energies together with
condition (\ref{fermivals}) to compute the field dependence of the
$\Lambda_r$ ($g_s^-(\omega)$ is given in Eq.~(\ref{defgpm})):
\begin{equation}
  \sin{\pi{r}\over N}\, \exp\left({-\pi\Lambda_r\over N}\right) =
     \sum_{s,t=1}^{N-1}
     \left( {{\sin\pi{}rs/N\, \sin\pi{}st/N}
            \over{\sin\pi{}s/2N\, g_s^-(-i\pi/N)}}\right)
            {\varepsilon_t^{(0)}\over \pi h_c}
  \label{fermival_sc}
\end{equation}
An analogous computation yields the actual field dependence of the
$Z_{rc}$ in Eq.~(\ref{zzch_sc}). For $h_s\ll h_c$ we find
\begin{equation}
   Z_{rc}={N-r \over N} - {N\over\pi^2}
      \sum_{s=1}^{N-1} s_<\left(N-s_>\right){h_s\over h_c}
   \label{zzch_sch}
\end{equation}
where $s_<$ ($s_>$) is the smaller (greater) of the integers $r$, $s$.

{}From (\ref{bareener}) the bare energies $\varepsilon_r^{(0)}$ are
known to be proportional to the applied magnetic fields, hence the
$\Lambda_r$ in (\ref{fermival_sc}) show the logarithmic dependence on
the fields found previously in the isotropic Heisenberg spin chain
\cite{bik86} and the SU(2) Hubbard model \cite{frko91}:
\begin{equation}
   \Lambda_r \sim \ln\left({h_c\over h}\right)
\end{equation}
where $h$ is the typical strength of the fields $h_r$.  As is known
from the SU(2) Hubbard model this strong dependence on small applied
fields shows up in the field dependence of the anomalous dimensions
since they contain terms $\propto 1/\Lambda_s$ as the leading
corrections in their spin-components $\Delta_r^\pm$ ($r=1,\ldots,N-1$)
(see Eq.~(\ref{dimsh0})) and is related to a nonanalytic field
dependence of the magnetic susceptibility in this system \cite{schl89}
and the \SUN{} Heisenberg model \cite{schl92b}.  This is in contrast
to the charge-components $\Delta_c^\pm$ of the conformal dimensions
where (\ref{zzch_sch}) implies a linear dependence on the applied
fields.

For sufficiently large fields ($h\simeq o(h_c)$) the system saturates in
a state with all electrons occupying states in the band(s) with the
lowest magnetic energy.  This final state depends on the particular
choice of the magnetic fields $h_s$ in (\ref{hamil}). In the following
we shall consider two possible cases explicitely, generalization to
others is straightforward.

%%%%%%%%%%%%%%%%%%%%%%%%%%%%%%%%%%%%%%%%%%%%%%%%%%%%%%%%%%%%%%%%%%%%%%
%
%
%\subsection*{Spin-S electrons}
%
%
%%%%%%%%%%%%%%%%%%%%%%%%%%%%%%%%%%%%%%%%%%%%%%%%%%%%%%%%%%%%%%%%%%%%%%
---One natural interpretation of the \SUN-index is that of an orbital
quantum number, i.e.\ taking the electrons in the $r$-th band as
having spin $S+1-r$ where $N=2S+1$.  In this picture the coupling to
the magnetic fields should be through Zeeman terms giving
\begin{equation}
   h_r = -(S+1-r)\,h, \qquad 1\le r \le N=2S+1
\end{equation}
for the fields in (\ref{hamil}) or
\begin{equation}
   e_r^{(0)} = h - {h_c\over{1+\lambda^2}} \delta_{r,1}.
\end{equation}
for the bare energies (\ref{bareen_sc}).  In this interpretation it is
straightforward to see that the quantity $h_c$ introduced in
Eq.~(\ref{bareen_sc}) is simply the large-$u$ limit of the critical
magnetic field beyond which the ground state of the system is
ferromagnetically ordered (i.e.\ all electrons are in the band with
spin-$S$).  For $h>h_c$ only excitations with $\Delta{}M_r=0$
($r=1\ldots,N-1$) are gapless. By construction, the corresponding
correlation functions are those of free spinless electrons.

%%%%%%%%%%%%%%%%%%%%%%%%%%%%%%%%%%%%%%%%%%%%%%%%%%%%%%%%%%%%%%%%%%%%%%
%
%
%\subsection*{Degenerate bands}
%
%
%%%%%%%%%%%%%%%%%%%%%%%%%%%%%%%%%%%%%%%%%%%%%%%%%%%%%%%%%%%%%%%%%%%%%%
---Another possible interpretation of the $N$ bands in the degenerate
Hubbard model is that of degenerate bands of spin-$\pm{1\over2}$
electrons.  To be specific let us choose the electrons in the first
$N_+$ bands as having spin $\uparrow$ and the ones in the remaining
$N_-=N-N_+$ bands as carrying spin $\downarrow$.  This choice gives
\begin{equation}
   h_r= -{h\over2} \quad {\rm for~} 0\le r \le N_+, \qquad
   h_r= {h\over2} \quad {\rm for~} N_+ < r \le N, \qquad
\end{equation}
for the coupling of a physical magnetic field to the system
corresponding to
\begin{equation}
   e_r^{(0)} = h \delta_{r,N_+} - {h_c\over{1+\lambda^2}} \delta_{r,1}.
\end{equation}
for the bare energies (\ref{bareen_sc}).  Again the system saturates
at large fields: for
\begin{equation}
   h\ge h_x = {h_c \over 2 N_+} \left( \psi\left({1\over2}+{1\over{N_+}}\right)
                                - \psi\left({1\over2}\right) \right)
\end{equation}
the groundstate of the system is determined by filled bands for the
$\uparrow$-electrons with densities $n_s=n_c/N_+$ while the $N_-$
bands of $\downarrow$-electrons are empty.  This state shows SU($N_+$)
spin-symmetry, excitations involving creation of electrons in one of
the $\downarrow$-bands are massive.  The gapless excitations and
corresponding critical exponents are given by the $u\to\infty$-limits
of the expressions in the preceeding section: the dressed charge
matrix for the critical degrees of freedom is of the form
(\ref{zmatrix}) with $N$ replaced by $N_+$ and $z_c=1$.

%% Date: Wed, 15 Jul 92 13:11:22 MET DST
%
\begin{appendix}
%
%%%%%%%%%%%%%%%%%%%%%%%%%%%%%%%%%%%%%%%%%%%%%%%%%%%%%%%%%%%%%%%%%%%%%%%%%%%
\section{Appendix}
In this appendix we briefly list some mathematical results which are helpful
in the solution of the Wiener-Hopf equations in Sections 4 and 5. The
Wiener-Hopf method itself has been reviewed in the appendix of \cite{frko91}.

In the reduction of the matrix Wiener-Hopf problem to scalar ones one has
to diagonalize a tridiagonal $(N-1)\times (N-1)$ Toeplitz-matrix of the type
\begin{equation}
T=\pmatrix{
    2y & -x     &  0     & \cdots & 0      \cr
    -x & 2y     & -x     & \ddots & 0      \cr
     0 & -x     & 2y     & \ddots & 0      \cr
 \vdots& \ddots & \ddots & \ddots & \vdots \cr
     0 & \cdots &        &   -x   & 2y\cr}
\label{toeplitz}
\end{equation}
where $x$ and $y$ are real numbers. In terms of $\lambda_\pm =y\pm \sqrt{y^2
-x^2}$ one finds for the determinant of $T$
\begin{equation}
%{\rm det}~T =\cases{{\lambda_+^N-\lambda_-^N \over \lambda_+-\lambda_-} & for
%                                          $\vert y\vert \ne \vert x\vert$ \cr
%               N   &for $\vert y\vert = \vert x\vert$ \cr}.
{\rm det}~T ={\lambda_+^N-\lambda_-^N \over \lambda_+-\lambda_-},
\label{det}
\end{equation}
where we assumed $\vert y\vert \ne \vert x\vert$.
Using this result the eigenvalues $t_j$ of $T$ are easily found to be
\begin{equation}
t_j=2\left( y-x\cos\left({j\pi \over N}\right)\right)
\label{toepevals}
\end{equation}
with corresponding eigenvectors
\begin{equation}
v^{(j)}=\pmatrix{v_1^{(j)} \cr
		 \vdots    \cr
		 v_{N-1}^{(j)} \cr}, \qquad
v_l^{(j)}=\sqrt{{2\over N}}\sin \left( {jl\pi \over N}\right).
\label{toepevecs}
\end{equation}
Thus $T$ is diagonalized by the matrix $U\in$\ O($N-1$) with elements
\begin{equation}
U_{jl}=\sqrt{{2\over N}}\sin \left( {jl\pi \over N}\right).
\label{uelements}
\end{equation}
The inverse of $T$ can be written as follows:
\begin{equation}
T^{-1}={1\over \det T}~S
\label{tinvers}
\end{equation}
where the elements of the matrix $S$ are given by
\begin{equation}
S_{jl} = a_{j_<}a_{N-j_>}x^{j_>-j_<}, \qquad
a_j    = {\lambda_+^j-\lambda_-^j \over \lambda_+-\lambda_-}.
\label{selements}
\end{equation}
Here again $j_<={\rm min\ }(j,l)$ and $j_>={\rm max\ }(j,l)$.

The main step of the Wiener-Hopf procedure is the factorization of the
Fourier-transformed kernel into a product of two functions $g^\pm$
which are analytic in the upper and lower complex $\omega$-plane,
respectively. In the derivation of (\ref{fermival_sc}) we have used
\begin{equation}
2e^{-\vert \omega\vert}\left( \cosh \omega -\cos{s\pi\over N}\right)
=g_s^+(\omega)g_s^-(\omega)
\label{kernel_fac}
\end{equation}
with
\begin{equation}
g_s^+(\omega)=g_s^-(-\omega)
=2\pi\left(-{i\omega\over\pi}\right)^{-{i\omega/\pi}}
{\exp\left(i{\omega\over\pi}\left(1+\ln 2\right)\right) \over
\Gamma\left({s\over 2N}-{i\omega\over 2\pi}\right)
\Gamma\left(1-{s\over 2N}-{i\omega\over 2\pi}\right)}
\label{defgpm}
\end{equation}
where $\Gamma$ denotes the Gamma-function. Note that the asymptotic behavior
of the functions defined in (\ref{defgpm}) is simply $\lim_{\omega\to\infty}
g_s^\pm(\omega) =1$.

Finally we want note that in the derivation of (\ref{znmatrix}) we do not
need the explicit form of the corresponding kernel $K(\omega)$ into
factors $G^\pm(\omega)$. One only needs the value $G^\pm(\omega =0)$ which is
-- due to the symmetry property $G^+(\omega)=G^-(-\omega)$ -- equal to
$\sqrt{G^+(0)G^-(0)}=\sqrt{K(0)}$ (see also \cite{bik86}).
\end{appendix}

%%Date: Fri, 3 Jul 92 11:46:08 MET DST
%

%%%%%%%%%%%%%%%%%%%%%%%%%%%%%%%%%%%%%%%%%%%%%%%%%%%%%%%%%%%%%%%%%%%%%%%%%%%
%
\newpage
\centerline{\large \bf Figure Captions}
\epsfysize=6cm
%%%%%%%%%%%%%%%%%%%%%%%%%%%%%%%%%%%%%%%%%%%%%%%%%%%%%%%%%%%%%%%%%%%%%%%%%%%
\begin{figure}[h]
\caption{\label{fig:1}}
\epsffile[-200 375 550 750]{Fig1.PS}
Dependence of the critical value $u_c$ the degeneracy $N$ as obtained from
(\ref{ucrit}). Note that for $u>u_c$ the system is in an insulating phase
whereas for $u<u_c$ it shows metallic behavior (for $n_c=1$).
\end{figure}
%%%%%%%%%%%%%%%%%%%%%%%%%%%%%%%%%%%%%%%%%%%%%%%%%%%%%%%%%%%%%%%%%%%%%%%%%%%
\epsfysize=11cm
\begin{figure}[h]
\caption{\label{fig:zcontours}}
\epsffile[50 0 850 480]{Fig2.PS}
%%\vspace{1 cm}
(a) Lines of constant $z_c$ for vanishing magnetic fields in the
$n_c$--$u$--plane for $N=2$. The drawn lines correspond to
$z_c=1.0904$, 1.1679, 1.2237, 1.2622, 1.2892, 1.3088, 1.3234
(corresponding to $z_0=0.4,0.8,\ldots,2.8$). Note that $z_c\to 1$ for
$u\to0$ and $z_c\to \sqrt{N}$ for $u\to\infty$ with $n_c$ arbitrary,
$n_c\to0$ with $u$ arbitrary, and $n_c\to1$ with $u>u_c$.  (b) As in
(a) but for $N=4$, $z_c=1.1241,1.2445,1.3473,1.4318,1.5013,
1.5586,1.6062$.  (c) As in (a) but for $N=6$,
$z_c=1.1316,1.2630,1.3799,1.4812,1.5689, 1.6453,1.7122$.  (d) As in
(a) but for $N=8$, $z_c=1.1344,1.2701,1.3928,1.5012,1.5973,
1.6830,1.7601$.
\end{figure}
%%%%%%%%%%%%%%%%%%%%%%%%%%%%%%%%%%%%%%%%%%%%%%%%%%%%%%%%%%%%%%%%%%%%%%%%%%%
\begin{figure}[h]
\caption{\label{fig:alpha}}
\epsffile[-220 0 550 750]{Fig3.PS}
Exponent $\alpha$ characterizing the Fermi-point singularity of the
momentum distribution function as function of the electron density
for $N=2,4,6,8,\infty$ (bottom to top) at zero field for two values
of $u$.
\end{figure}


\begin{thebibliography}{99}
%
%%%%%%%%%%%%%%%%%%%%%%%%%%%%%%%%%%%%%%%%%%%%%%%%%%%%%%%%%%%%%%%%%%%%%%%%%%%

\bibitem{hald8s}
  F. D. M. Haldane, Phys.\ Rev.\ Lett.\ {\bf 45}, 1358 (1980);
  J. Phys.\ C{\bf14}, 2589 (1981).

\bibitem{bpz}
  A. A. Belavin, A. M. Polyakov and A. B. Zamolodchikov,
  Nucl.\ Phys.\ B{\bf 241}, 333 (1984).

\bibitem{card86}
  J. L. Cardy, Nucl.\ Phys.\ B{\bf 270} [FS{\bf 16}], 186 (1986).

\bibitem{bcn86}
  H. W. J. Bl\"ote, J. L. Cardy and M. P. Nightingale,
  Phys.\ Rev.\ Lett.\ {\bf 56}, 742 (1986);
  I. Affleck, {\it ibid.} {\bf 56}, 746 (1986).

\bibitem{ikr89}
  A. G. Izergin, V. E. Korepin, and N. Yu.\ Reshetikhin,
  J. Phys.\ A{\bf 22},2615 (1989).

\bibitem{fryu90}
  H. Frahm and N.-C.\ Yu, J. Phys.\ A{\bf 23}, 2115 (1990).

\bibitem{woyn89}
  F. Woynarovich, J. Phys.\ A{\bf 22}, 4243 (1989).

\bibitem{frko90}
  H. Frahm and V. E. Korepin, Phys.\ Rev.\ B{\bf 42}, 10553 (1990).

\bibitem{frko91}
  H. Frahm and V. E. Korepin, Phys.\ Rev.\ B{\bf 43}, 5653 (1991).

\bibitem{kaya91}
  N. Kawakami and S.-K.\ Yang,
  J. Phys.: Condens.\ Matt.\ {\bf 3}, 5983 (1991).

\bibitem{ogat91}
  M. Ogata, M. U. Luchini, S. Sorella, and F. F. Assaad,
  Phys.\ Rev.\ Lett.\ {\bf 66}, 2388 (1991).

\bibitem{yang67}
  C. N. Yang, Phys.\ Rev.\ Lett.\ {\bf 19}, 1312 (1967).

\bibitem{hald80}
  F. D. M. Haldane, Phys.\ Lett.\ {\bf 80}A, 281 (1980); {\it ibid.\ }
  {\bf 81}A, 545 (1981) erratum.

\bibitem{schl91}
  P. Schlottmann, Phys.\ Rev.\ B{\bf 43}, 3101 (1991).

\bibitem{as91}
  A. Schadschneider, Dissertation, Universit\"at zu K\"oln (1991).

\bibitem{choy80}
  T. C. Choy, Phys.\ Lett.\ {\bf 80}A, 49 (1980).

\bibitem{chha82}
  T. C. Choy and F. D. M. Haldane, Phys.\ Lett.\ {\bf 90}A, 83 (1982).

\bibitem{schl89}
  K. Lee and P. Schlottmann, Phys.\ Rev.\ Lett.\ {\bf 63}, 2299 (1989).

\bibitem{schl90}
  K. Lee and P. Schlottmann, Physica B{\bf 163}, 398 (1990).

\bibitem{schl92}
  P. Schlottmann, Phys.\ Rev.\ B{\bf 45}, 5784 (1992).

\bibitem{lw68}
  E. H. Lieb and F. Y. Wu, Phys.\ Rev.\ Lett.\ {\bf 20}, 1445 (1968).

\bibitem{ksz90}
  A. Kl\"umper, A. Schadschneider and J. Zittartz, Z.\ Phys.\ {\bf B 78}, 99
  (1990).

\bibitem{sz91}
  A. Schadschneider and J. Zittartz, Z.\ Phys.\ {\bf B 82}, 387 (1991).

\bibitem{su75}
  B. Sutherland, Phys.\ Rev.\ B{\bf 12}, 3795 (1975).

\bibitem{yaya66}
  C. N. Yang and C. P. Yang, Phys.\ Rev.\ {\bf 150}, 327 (1966).

\bibitem{vega88}
  H. J. de Vega, J. Phys.\ A{\bf 21}, L1089 (1988).

\bibitem{suzu88}
  J. Suzuki, J. Phys.\ A{\bf 21}, L1175 (1988).

\bibitem{bik86}
  N. M. Bogoliubov, A. G. Izergin and V. E. Korepin, Nucl.\ Phys.\
  {\bf B275} [FS17], 687 (1986).

\bibitem{schl92b}
  P. Schlottmann, Phys.\ Rev.\ B{\bf 45}, 5293 (1992).

\end{thebibliography}
\end{document}